\documentstyle[prl,aps,twocolumn,epsf]{revtex}
\begin{document}

\title{Double-exchange via degenerate orbitals}

\author{Jeroen van den Brink}
\address{
Max-Planck-Institut f\"ur Festk\"orperforschung,
Heisenbergstrasse 1, D-70569 Stuttgart, Germany }
\author{Daniel Khomskii}
\address{
Laboratory of Applied and Solid State Physics, Materials Science Center,
University of Groningen, Nijenborgh 4, 9747 AG Groningen, The Netherlands}
\date{\today}
\maketitle
\begin{abstract}
We consider the double-exchange  for systems in which doped electrons occupy
degenerate orbitals,
treating the realistic situation with double degenerate $e_g$
orbitals.
We show that the orbital degeneracy leads in general to
formation of anisotropic magnetic structures and that
in particular, depending on the doping concentration, the layered magnetic structures
of the A-type and chain-like structures of the C-type are stabilized.
The phase-diagram that we obtain provides an explanation for the experimentally observed
magnetic structures of some over-doped (electron-doped) manganites of the type
Nd$_{1-x}$Sr$_x$MnO$_3$, Pr$_{1-x}$Sr$_x$MnO$_3$ and
Sm$_{1-x}$Ca$_x$MnO$_3$ with $x > 0.5$.
\end{abstract}

The double-exchange (DE) model~\cite{Zener51,Hasegawa55,DeGennes60} is one of the main models of
ferromagnetism in metallic systems. The interest in this model was renewed recently in
connection with the active study of the manganites with colossal
magneto-resistance (CMR)~\cite{Schiffer95}. Although it is not yet clear whether DE alone can
explain the CMR in manganites~\cite{Millis95}, and there are some problems with the model itself
(e.g. the tendency to phase separation~\cite{Nagaev69,Arovas,Kagan,Dagotto98}),
the double-exchange mechanism is a necessary ingredient for the theoretical understanding
of the CMR-effect~\cite{Millis95,Varma96} and remains the main explanation at least of the
presence of the ferromagnetic state in doped manganites of the type La$_{1-x}$M$_x$MnO$_3$ (M=Ca,Sr).

Most experimental data in the study of the CMR-materials are obtained
for the hole-doped manganites ($x < 0.5$) as it is usually in this
doping range that CMR is observed. Recently, however, interesting data appeared
on the properties of over-doped (or electron-doped)
manganites ($x > 0.5$). Theoretically, in a conventional DE model one expects
qualitatively similar behavior for small $x$ and for $x \sim 1$.
Experimentally, however, there exists a very strong asymmetry. The
behavior of doped manganites for $x < 0.5$ and $x > 0.5$ is
drastically different. In particular a very stable insulating stripe
phase is observed for $x > 0.5$ with the period determined by
doping~\cite{Cheong}, whereas for $x < 0.5$, depending on the exact
composition, either the metallic ferromagnetic state is
realized~\cite{Schiffer95}, or a charged-ordered state with a doubled
unit cell~\cite{Jirac85}, which, however, can be easily transformed
into a ferromagnetic metal by a magnetic field~\cite{Tokura}. Recently
Maignan {\it et al.}~\cite{Maignan98,Raveau} succeeded in obtaining
the state similar in some respects to the one predicted by the DE
model in the electron-doped region.  In Ca$_{1-y}$Sm$_y$MnO$_3$ they
observe metallic-like behavior with unsaturated ferromagnetism for
$y=7-12\%$ with indications that the state is not really
ferromagnetic but has a more complicated, possibly canted, magnetic
order.

Other recent data~\cite{Akimoto98,Kuwahara,Kawano97,TokuraP} show that the 
system Nd$_{1-x}$Sr$_x$MnO$_3$ is ferromagnetic for $0.25< x < 0.5$,
A-type antiferromagnetic ordered for $0.5 < x < 0.6$ and that
C-type antiferromagnetic order is realized for
$0.6<x<0.8$. In Pr$_{1-x}$Sr$_x$MnO$_3$ a A-type magnetic structure
exists for $0.5<x<0.7$. All these results show again that the behavior
of the hole-doped and electron-doped manganites is indeed very
different.

There is an important factor which is overlooked until now when one tries to apply the DE
picture to these transition metal oxides.
The $e_g$ levels, or bands, in which the charge carriers move are orbitally degenerate.
The first attempt to take this into account is the
recent work of Gor'kov and Kresin~\cite{Gorkov98}, who aimed to explain, starting from
a band-picture, the properties of undoped and slightly doped LaMnO$_3$.
They show in particular that one can indeed rationalize in this way some properties of
La$_{1-x}$M$_x$MnO$_3$ for small $x$. They neglect, however, the strong Coulomb
interactions that make undoped LaMnO$_3$ a Mott-Hubbard insulator, making their
approach in this regime somewhat questionable.

In undoped LaMnO$_3$ the localized $e_g$ electrons are orbitally ordered, lifting the
orbital degeneracy. Thus for the doped system we can in a first approximation
ignore the orbital degeneracy and apply the conventional DE model. If we however
proceed from the opposite direction and consider e.g. adding electrons to the insulating
CaMnO$_3$, containing Mn$^{4+}$ ($t^3_{2g}$) ions with localized moments in the
$t_{2g}$ levels, one dopes these electrons into the empty $e_g$
levels, which are orbitally degenerate. Thus we have now a problem of
{\it double-exchange via degenerate orbitals}. As we show below, this
orbital degeneracy drastically modifies the picture of the DE and
results in the formation of anisotropic magnetic structures. Our
theoretical results have much in common with the behavior of
manganites in this doping range as found in
experiments~\cite{Maignan98,Raveau,Akimoto98,Kuwahara,Kawano97,TokuraP}.

We start with the model that is realistic, for example, for Ca$_{1-y}$R$_y$MnO$_3$
(R is a rare earth, $y$ is small). For $y=0$ this system is a Mott insulator
with localized $t_{2g}$-electrons (the localized spin {\bf S}) with  antiferromagnetic
exchange interaction $J {\bf S}_i \cdot {\bf S}_j$, where $i$ and $j$ are nearest
neighbors, leading to a G-type (simple two-sublattice) antiferromagnetic ground-state.
Doped electrons enter the doubly degenerate $e_g$-levels. The strong on-site Hund's rule
exchange tends to align the spins of the $t_{2g}$ and $e_g$ electrons parallel.
The effective Hamiltonian describing the low energy properties for this system is thus
\begin{eqnarray}
H&=&J \sum_{<ij>}{\bf S}_i \cdot {\bf S}_j
 - J_H \sum_{i,\alpha,\sigma,\sigma{^\prime}} {\bf S}_i \cdot c_{i \alpha \sigma}^{\dag}
       \vec{\sigma}_{\sigma \sigma{^\prime}} c_{i \alpha \sigma{^\prime}} \nonumber \\
 &-&\sum_{<ij>,\sigma} t_{ij}^{\alpha \beta}
                       c_{i \alpha \sigma}^{\dag} c_{j \beta \sigma}.
\label{eq:H}
\end{eqnarray}
The first and the second term are explained above, and the last term describes the hopping
of $e_g$-electrons. These electrons are labeled by the site index $i$, spin $\sigma$
and orbital index $\alpha (\beta)=1,2$, corresponding e.g. to the $d_{x^2-y^2}$ and
$d_{3z^2-r^2}$ orbitals. The presence of the orbital degeneracy, together with the
very particular relations between the hopping matrix elements
$t_{ij}^{\alpha \beta}$~\cite{Anderson59,Kugel82}
makes this problem, and its outcome, very different from the usual DE case.

Our approach follows the standard route~\cite{DeGennes60} of treating
the spin subsystem quasi-classically and assuming a homogeneous ground-state.
Furthermore we assume that the intra-atomic Hund's rule exchange
interaction $J_H$ is strong, so that only states with the maximal possible spin
at each site are realized and that the inter-site exchange interaction
$J$ is smaller that the electron hopping integral $t$. Both are certainly good
assumptions for the manganites.

In a first step we have to determine the spectrum of the
$e_g$-electrons ignoring their interaction with the localized spins.
This spectrum is given by the solution of the matrix equation
\begin{eqnarray}
|| t_{\alpha \beta}- \epsilon \delta_{\alpha \beta} || =0,
\label{eq:ME}
\end{eqnarray}
where
\begin{eqnarray}
t_{11}&=& -2t_{xy}(\cos k_x + \cos k_y) \nonumber \\
t_{12}&=& t_{21}= -\frac{2}{\sqrt{3}} t_{xy}(\cos k_x - \cos k_y) \nonumber \\
t_{22}&=& -\frac{2}{3} t_{xy}(\cos k_x + \cos k_y)
         -\frac{8}{3} t_z \cos k_z.
\label{eq:hop}
\end{eqnarray}
Here $t_{11}$ is the dispersion due to an overlap between
$d_{x^2-y^2}$-orbitals on neighboring sites, $t_{12}$ --- between a
$d_{x^2-y^2}$ and $d_{3z^2-r^2}$ orbital and $t_{22}$ between two
$d_{3z^2-r^2}$ orbitals. In writing Eq.~(\ref{eq:hop}) we have taken
into account the ratios of different hopping
integrals~\cite{Anderson59,Kugel82}, which are determined by the
symmetry of the $e_g$ wavefunctions, and introduced the notation
$t_{xy}$ and $t_z$, to be defined furtheron. In absence of magnetic
order, $t_{xy}=t_z=t$.  The solutions $\epsilon_{\pm}( {\bf k} )$ of
Eq.~(\ref{eq:ME}) are
\begin{eqnarray}
\epsilon_{\pm}( {\bf k} )
&=&  -\frac{4t_{xy}}{3} (\cos k_x + \cos k_y)   -\frac{4t_z}{3}\cos k_z \nonumber \\
&\pm&
 \left(
      \left[ \frac{2t_{xy}}{3}  (\cos k_x + \cos k_y)  -\frac{4t_z}{3}\cos k_z \right]^2
\right.    \nonumber \\
&+&  \left. \frac{4t_{xy}^2}{3} (\cos k_x - \cos k_y)^2 \right) ^{1/2}
\label{eq:disp}
\end{eqnarray}
These bands agree with those obtained in Ref.~\cite{Gorkov98}.

We now take into account the interaction of the $e_g$-electrons with a magnetic
background. We assume that the underlying magnetic structure is characterized
by two sub-lattices, with a possible canting, so that the angle between neighboring
spins in the $xy$-plane is $\theta_{xy}$ and in the $z$-direction is  $\theta_{z}$.
This rather general assumption covers the G-type
antiferromagnets (two-sub-lattice structure,
$\theta_{xy}=\theta_{z}=\pi$), A-type one (ferromagnetic planes
coupled antiferromagnetically, $\theta_{xy}=0, \ \theta_{z}=\pi$) and
C-type structures (antiferromagnetically coupled ferromagnetic chains,
$\theta_{xy}=\pi, \ \theta_{z}=0$).  For angles different from these,
we  get a canted spin structure which originates from a simple
antiferromagnetic one (G-, A-, or C-type).
Following~\cite{DeGennes60} we then have the effective hopping matrix
elements $t_{xy}=t \cos (\theta_{xy}/2)$ and $t_z=t \cos
(\theta_z/2)$.  This clarifies the notation introduced in
Eq.~(\ref{eq:hop}). As in the quasi-classical treatment of the
non-degenerate DE model, the energy spectrum~(\ref{eq:disp}) is
renormalized by the magnetic order and, because of the
orbital-dependent hopping matrix elements in a degenerate system, this
results, in general, in an anisotropic magnetic structure.

When we dope the system with electrons, these go into states with minimal energy,
in our case into the $\Gamma$-point ${\bf k}=0$.
Let us first assume that all doped charges go into the state with
lowest energy, which is strictly speaking only the case for very small doping,
and neglect for the moment the effects due to a finite filling of the bands.
At the $\Gamma$-point the energies are
$\epsilon_{\pm}(0) = -\frac{4}{3}(2t_{xy}+t_z)\pm\frac{4}{3} | t_{xy}-t_z |$.
We can now write down the total energy per site of our system
containing $y$ electrons ($y=1-x$ for conventional notation for e.g.\
La$_{1-x}$A$_x$MnO$_3$)
\begin{eqnarray}
E&=&\frac{J}{2} (\cos \theta_z + 2\cos \theta_{xy})+ y \epsilon_{min}(0) \nonumber \\
 &=& -\frac{3J}{2} + 2J \cos^2 (\theta_{xy}/2) +J\cos^2 (\theta_z/2)
     \nonumber \\
 &-&\frac{4}{3} y t    \left[ \right. 2 \cos (\theta_{xy}/2) + \cos (\theta_z/2) \nonumber \\
 &+&  | \cos (\theta_{xy}/2) - \cos (\theta_z/2) | \left. \right].
\label{eq:energy}
\end{eqnarray}

There exist two possible situations.
If $\cos (\theta_{xy}/2) > \cos (\theta_z/2)$, then the magnetic structure is
A-type-like; in this case the minimization of the energy with respect
to the angles $\theta_{xy}$ and $\theta_z$ gives
\begin{eqnarray}
\cos (\theta_{xy}/2) = \frac{t}{J}y, \ \ \ \cos (\theta_z/2) =0
\label{eq:angle1}
\end{eqnarray}
and the energy of the corresponding state amounts to
\begin{eqnarray}
E^{(1)} = -\frac{3}{2}J-\frac{2t^2}{J} y^2
\label{eq:energy1}
\end{eqnarray}
Physically this state corresponds to an $xy$-plane with a canted 
structure, with the spins in neighboring planes being antiparallel.
If $\cos (\theta_{xy}/2) < \cos (\theta_z/2)$, then the magnetic structure is
C-type-like and one finds for
$\theta_{xy}$ and $\theta_z$
\begin{eqnarray}
\cos (\theta_{xy}/2) = \frac{t}{3J}y, \ \ \ \cos (\theta_z/2) =\frac{4t}{3J}y.
\label{eq:angle2}
\end{eqnarray}
The energy of this state $E^{(2)}$ is exactly equal to that of the
A-type state $E^{(1)}$.
In this situation we have a canted structure in all three directions with the
correlations in the $z$-direction being ``more ferromagnetic''.
Thus in the lowest-order in $J$ approximation the two
solutions~(\ref{eq:angle1}) and~(\ref{eq:angle2}) are degenerate.

One can easily show that in this approximation (filling of only ${\bf
k}=0$ point and ignoring the Fermi-energy for finite concentration) we
would have degenerate solutions up to a concentration
$y_c=\frac{3J}{4t}$ beyond which the A-type solution becomes more
favorable.  It is also worthwhile to note that this solution never
evolves into a ferromagnetic one: for $y>J/t$ the canting angle
$\theta_{xy}$ in Eq.~(\ref{eq:angle1}) is zero and the basal plane becomes
ferromagnetic, but the moments of the neighboring planes will be
opposite to it, i.e.\ we obtain pure A-type antiferromagnetism.  In
contrast to that, the C-type-like solution~(\ref{eq:angle2}) would
give with increasing $y$ first the state with completely ferromagnetic
chains but with neighboring chains at certain angle to those and
finally, at $y=3J/t$, also the angle $\theta_{xy}=0$ and we would get
an isotropic ferromagnetic state.  However the A-type solution has
lower energy in this region; thus we would never reach ferromagnetic
state in this approximation, contrary to experiment.

The stability of the A-phase is easy to explain: in this state the
$d$-electrons occupy predominantly the orbitals of
$d_{x^2-y^2}$-character and their delocalization requires the spins in
$xy$ plane to be parallel, while 
spin correlations in $z$-direction may remain antiferromagnetic.
However this all is true only at the $\Gamma$-point (${\bf k}=0$).
With increasing electron doping the higher lying states will be filled
which will modify the picture.

We therefore perform the same calculation
as above, but taking into account that at larger doping concentrations the $e_g$ bands
gradually fill up. In order to do so, we fix the doping concentration $y$ as well as the
ratio $t/J$
and numerically evaluate the values of $\theta_{xy}$ and $\theta_z$ which minimize
the total energy, taking into account that for every pair of angles
the bands given in Eq.~(\ref{eq:disp}) are filled up to the
Fermi-energy.

We find that at very low doping concentrations the A-type solution given
by Eq.~(\ref{eq:angle1}) and the C-type solution given by
Eq.~(\ref{eq:angle2}) are indeed the magnetic structures of lowest
energy, in accordance with our analytical treatment.  We find,
however, that in this region the A-type solution is always slightly
lower in energy than the C-type solution, i.e.\ the degeneracy of
these two states is lifted. This has a simple physical reason. In the
A-type structure the dispersion of the bands is strictly
two-dimensional (2D), so that the density of states (DOS) at the band
edge is non-zero.  For the C-type solution of Eq.~(\ref{eq:angle2}),
the bands have a highly anisotropic but already three-dimensional
character, leading typically to a vanishing DOS at the band edge.
Therefore the A-type magnetic structure is stabilized as it has a
larger DOS at the band-edge. At a somewhat higher doping level, however,
the quasi-one-dimensional peak in the DOS close to the band-edge
starts to play a role and can cause the transition to a C-type state.

\begin{figure}
      \epsfysize=50mm
      \centerline{\epsffile{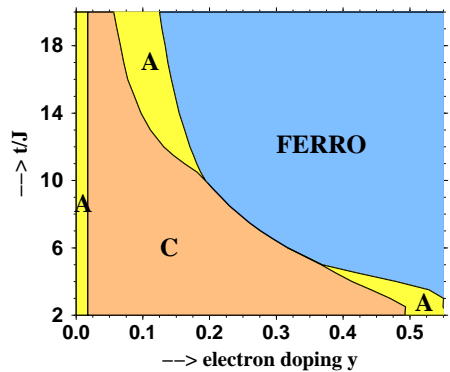}}
\caption
{Phase diagram of the double-exchange model with degenerate $e_g$ bands. Depending
on the electron doping concentration and the ratio of the $e_g$ bandwidth and the $t_{2g}$
superexchange one finds A-type, C-type or ferromagnetic magnetic order.}
\label{fig:phase}
\end{figure}

The full phase-diagram resulting from our calculations is presented in
Fig.~\ref{fig:phase}. At very low electron doping we find that the
A-phase  (with canting) is stable for all values of $t/J$. Increasing
the doping, the system first enters the C-phase (also canted), and,
depending on the value of $t/J$, re-enters the A-phase before becoming
ferromagnetic at large electron doping.  
This sequence of phases is determined by the form of the DOS of this
structures. 
Our phase-diagram has certain features in common with that
of Maezono {\it et al.}~\cite{Maezono98}, although our approach is quite different.

We now compare our results with experimental data on electron doped manganites.
For example, for $t/J \approx 4$, we find a stable ferromagnetic phase
up to $x \approx 0.5$ (our $y=1-x$) and the A-phase up to $x \approx
0.6$, and for larger hole doping we find the canted C-type phase to be
stable, in agreement with observations in
Nd$_{1-x}$Sr$_x$MnO$_3$~\cite{Akimoto98,Kuwahara}.  The A-type state also
strongly resembles the one observed in
Pr$_{1-x}$Sr$_x$MnO$_3$~\cite{Kawano97,TokuraP}.  Moreover, there are
indications that the unsaturated ferromagnetic and (semi-)metallic
state observed for Ca$_{1-y}$R$_y$MnO$_3$ in Ref.~\cite{Maignan98,Raveau} 
also has C-type magnetic correlations~\cite{Mahendiran}. This is consistent
with the fact that our C-type solution can have canting in all three
directions and consequently has a net magnetic moment, whereas the
A-type solution has no net moment.

Our calculations show that in the C-type phase the
electronic band is nearly one-dimensional.  In this case any disorder
renders the 1D system insulating. This provides a possible explanation
for the observation that the C-type phase is insulating for electron
doping above $y=0.15$~\cite{Kuwahara}.  According to the same
argument, the 2D bands of A-type phase and the 3D dispersion in
ferromagnetic phase makes these magnetic structures less susceptible
to disorder so that they are more likely to show metallic behavior, as
observed in experiment.

We note also that there exist other physical factors that may 
stabilize 
a particular state and that can therefore modify details of the
phase-diagram of Fig.~\ref{fig:phase}.  In the A-type solution, for
instance, predominantly the $x^2-y^2$ orbitals are
occupied, giving rise to a lattice distortion --- in this case a
compression of the MnO$_6$ octahedra along the $z$-direction. One can
show, however, that due to the anharmonicity the Jahn-Teller
distortion always leads to a local {\it elongation}~\cite{Anisimov}.
If strong enough, this tendency would favor a 
structure in which the $3z^2-r^2$ orbitals are occupied, in this case
lowering the energy of C-type structures.

As already mentioned in the introduction, there are some subtle points in the
conventional treatment of the DE model, as adapted above. All these
complications (quantum nature of the spins, possible existence of inhomogeneous
states) can be present also in the degenerate case. 
For large doping, one should also consider the interaction between $e_g$
electrons themselves.
Nevertheless we believe that the
main consequences of the degeneracy of electronic states responsible
for the double exchange, e.g. in over-doped manganites, are correctly 
captured in our treatment.

We conclude that the double-exchange via degenerate orbitals has very interesting special
features that are very different from the conventional, non-degenerate, situation.
In particular we have shown that the double-exchange via $e_g$ states naturally leads to
formation of anisotropic magnetic structures, layered or chain-like, in otherwise cubic compounds.
We have presented the phase-diagram for this model.
Our treatment has a general applicability because orbital degeneracy is experimentally
found in many real systems. In particular it can serve as a basis for an explanation
of magnetic and transport properties of over-doped (electron-doped) manganites.

We are grateful to O. Jepsen for making computer routines available 
to us, to B. Raveau, Y. Tokura and R. Mahendiran for informing us of some 
experimental results prior to publication, and especially to L.P. 
Gor'kov for very useful discussions.
J.v.d.B. acknowledges with appreciation the support by the Alexander
von Humboldt-Stiftung, Germany.
This work was financially supported by the Nederlandse Stichting voor Fundamenteel
Onderzoek der Materie (FOM) and by the European network OXSEN.

\end{document}